\documentclass[twocolumn,english,aps,prl,showpacs]{revtex4}
\usepackage[T1]{fontenc}
\usepackage{amsmath,graphicx,amssymb,epsfig,babel,dsfont}

\newcommand{\Tr}{{\rm Tr}}

\newcommand{\rL}{{\rm L}}
\newcommand{\rR}{{\rm R}}
\newcommand{\ri}{{\rm i}}

\begin{document}

\title{Phase-change radiative thermal diode}

\author{Philippe Ben-Abdallah}
\email{pba@institutoptique.fr}
\affiliation{Laboratoire Charles Fabry,UMR 8501, Institut d'Optique, CNRS, Universit\'{e} Paris-Sud 11,
2, Avenue Augustin Fresnel, 91127 Palaiseau Cedex, France.}
\author{Svend-Age Biehs}
\email{s.age.biehs@uni-oldenburg.de}
\affiliation{Institut f\"{u}r Physik, Carl von Ossietzky Universit\"{a}t, D-26111 Oldenburg, Germany.}

\date{\today}


\begin{abstract}
A thermal diode transports heat mainly in one preferential direction rather than in the 
opposite direction. This behavior is generally due to the non-linear dependence of certain 
physical properties with respect to the temperature. Here we introduce a radiative thermal 
diode which rectifies heat transport thanks to the phase transitions of materials. 
Rectification coefficients greater than 70\% and up to 90\% are shown, even for small 
temperature differences. This result could have important applications in the development 
of futur contactless thermal circuits or in the conception of radiative coatings for 
thermal management.  
\end{abstract}

\maketitle

Asymmetry of heat transport with respect to the sign of the temperature gradient between two points 
is the basic definition of thermal rectification~\cite{Starr1936,RobertsWalker2011} which is 
at the heart of a variety of applications as for example in thermal regulation, thermal  modulation, 
and heat engines. This unusual thermal behavior has opened the way to new concepts for manipulating 
the heat flow similarly to the electric current in electronic devices. Usually this manipulation finds its 
origin in the non-linear behavior of materials with respect to the temperature,  which, for the thermal 
rectification, breaks the symmetry of transfer when the temperature gradient is reversed. The 
effectiveness of the thermal rectification can be measured by means of the rectification 
coefficient $\eta =\frac{\mid\Phi_F-\Phi_R\mid}{\max(\Phi_F;\Phi_R)}$, where $\Phi_F$ and  $\Phi_R$ denote 
the heat flux in the forward and reverse operating mode, respectively. Different solid-state thermal 
diodes have been conceived during the last decade from various mechanisms (see \cite{BaowenLiEtAl2012} for a review on 
phononic rectification), including nonlinear atomic vibrations~\cite{Casati}, nonlinearity of the 
electron gas dispersion relation in metals~\cite{Segal}, direction dependent 
Kapitza resistances~\cite{Gong} or dependence of the superconducting density of states and phase dependence of 
heat currents flowing through Josephson junctions~\cite{Giazotto}. 

\begin{figure}[Hhbt]
\includegraphics[scale=0.35]{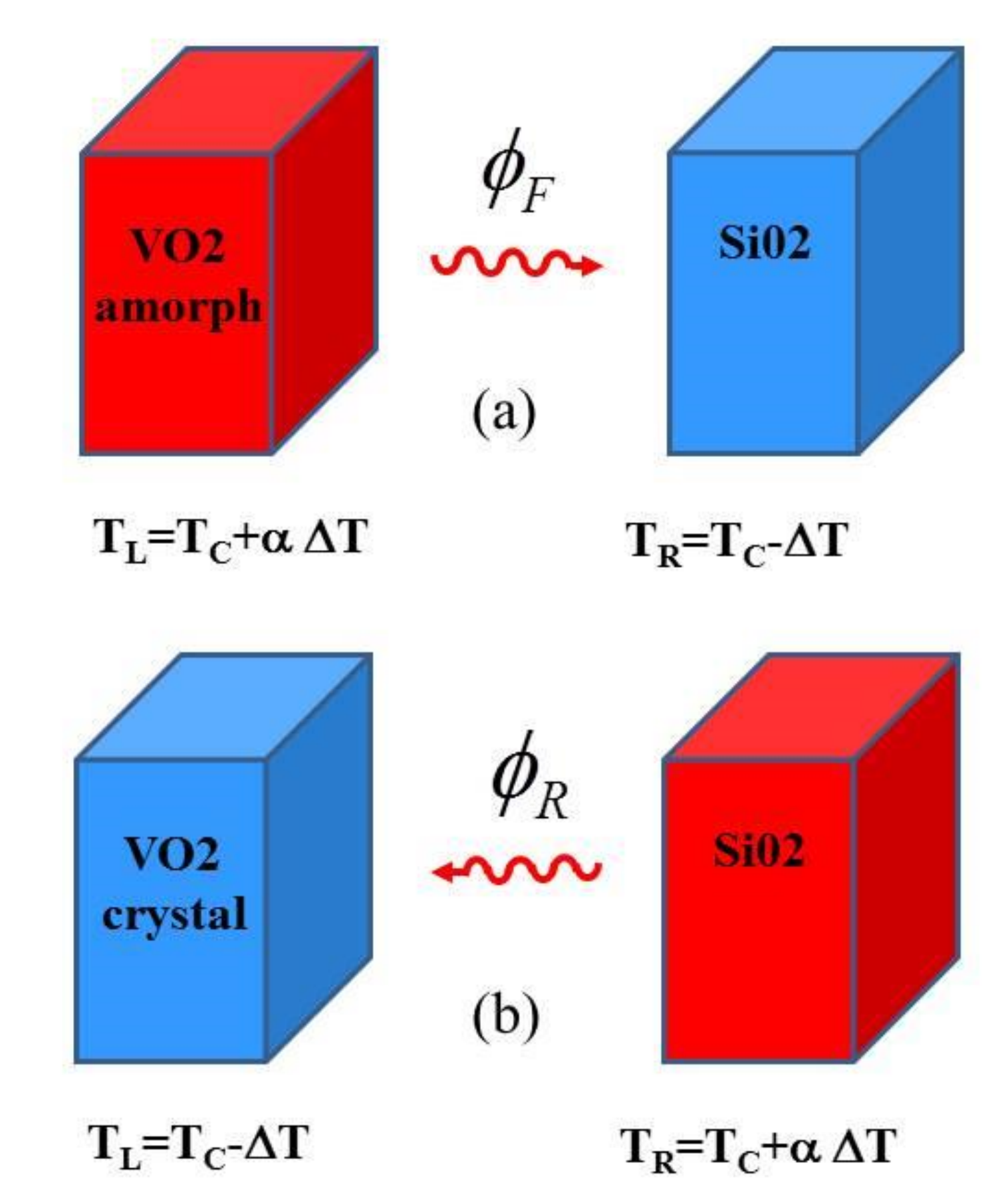}

\caption{Schematic of a phase-change radiative thermal diode. (a) Forwad scenario: the phase-change 
         material is its amorph state at higher temperature than its transition temperature$ T_c$. 
         (b) Reverse scenario: the phase change material is in its cristalline state. 
         \label{Fig:Sketch}}
\end{figure}

More Recently, photon-mediated thermal rectifiers ~\cite{OteyEtAl2010,IizukaFan2012} have been proposed 
to tune near-field heat exchange using materials with thermally dependent optical resonances. Since then, 
numerous mechanisms have been introduced to manipulate the non-radiative heat 
exchanges \cite{BiehsEtAl2011a,VanZwolEtAl2011,BasuFrancoeur2011,ZhuEtAl2012} between two bodies. 
Recently a far-field thermal rectifier has been proposed on the 
basis of spectrally selective micro or nanostructured thermal emitters~\cite{NefzaouiEtAl2013}  as previously developed 
to design coherent thermal sources~\cite{DrevillonEtAl2011} and to enhance the near-field thermal emission of composite 
structures \cite{pba}. However, so far only relatively weak radiative and non-radiative thermal 
rectifications have been highlighted with these mechanisms ($\eta< 44\%$ in Refs.~\cite{OteyEtAl2010,IizukaFan2012}, $\eta < 52\%$ in Ref.~\cite{BasuFrancoeur2011}, and  $\eta< 70\%$ in Ref.~\cite{NefzaouiEtAl2013} for instance).

In this Letter, we propose a radiative thermal rectification principle based on the phase transition  of insulator-metal transition (IMT) materials around the operating temperature.  In an IMT material a small smooth change of the temperature around its critical temperature $T_c$ causes a sudden qualitative and quantitative change (i.e. a bifurcation) in its optical properties \cite{Baker}. In a recent work van Zwol {\itshape et al.} 
have shown that the near-field heat-flux exchanged between two media, at close separation distances 
(subwavelength), could be modulated by several orders of magnitude across the phase transition of 
vanadium dioxide (VO$_2$) deposited on one of surfaces~\cite{VanZwolEtAl2011}. Here the concept of a far-field radiative 
thermal rectifier is presented. We show that the bifurcation of optical properties across the phase 
transition between the amorphous and the crystalline phases of VO$_2$ can lead to a thermal rectification 
coefficient which is greater than 70\% for small temperature differences and can even be larger 
than 90\% making these devices serious candidates for thermal diodes without any contact 
between the high and the low temperature regions

\begin{figure}[Hhbt]
\includegraphics[scale=0.35]{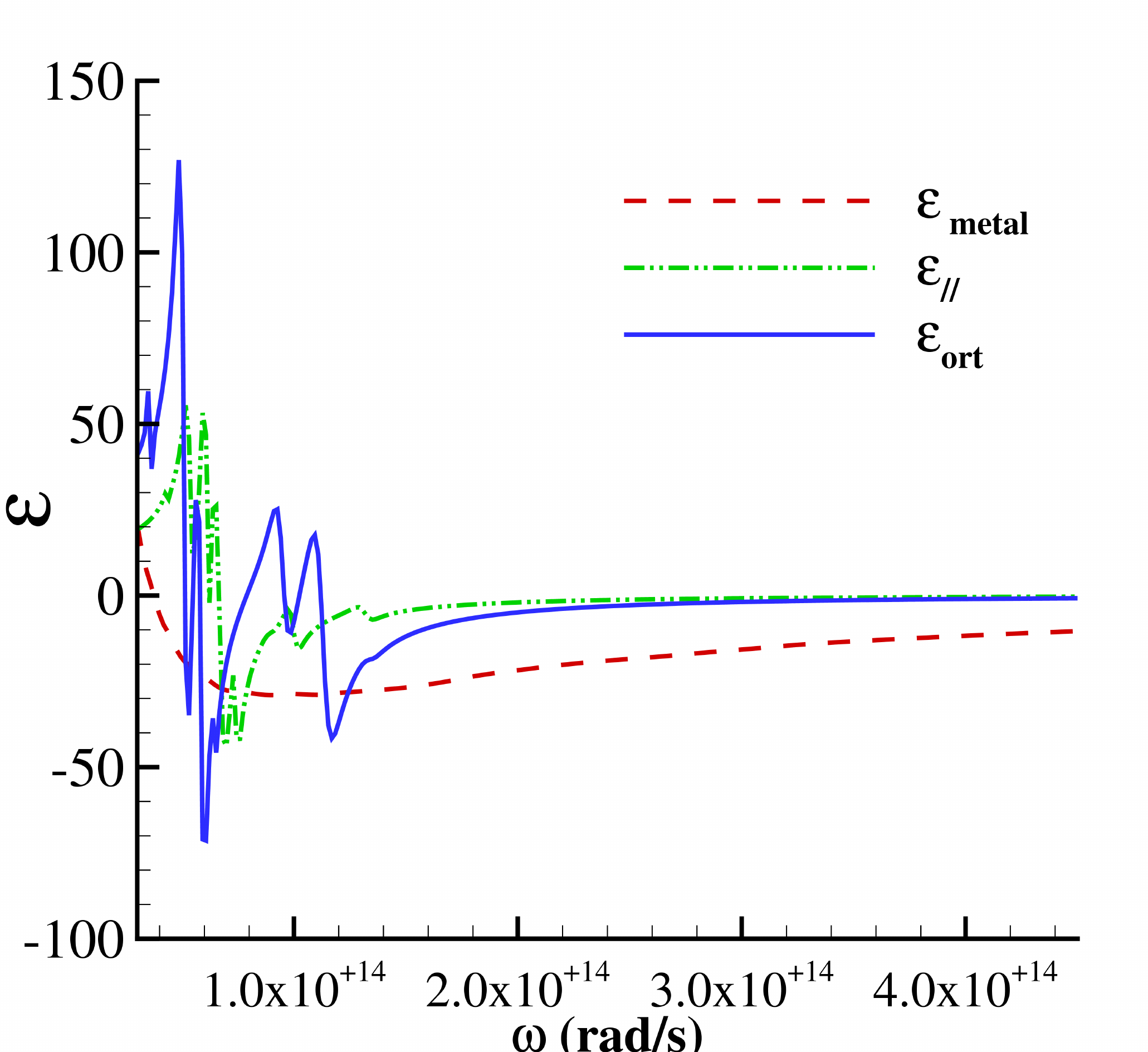}
\caption{ Permittivities of VO$_2$ in the metallic phase ($\epsilon_{\rm metal}$) and in the cristalline 
          phase ($\epsilon_\parallel$ and $\epsilon_\perp$). \label{Fig:Perm}}
\end{figure}

To start, let us consider a system as illustrated in Fig.~\ref{Fig:Sketch}, where two semi-infinite plane 
bodies, one made of VO$_2$ and one made of amorphous glass (Si0$_2$) at temperatures $T_\rL$ and $T_\rR$, 
respectively. Both media are separated by a vacuum gap of thickness $d$ which is assumed to be much 
larger than their thermal wavevelengths so that heat exchanges are mainly due to propagating photons,
i.e.\ we are considering the far-field regime. We now examine this system in the two following thermal 
operating modes: (i) In the forward mode (F) the temperature $T_\rL = T_c+\alpha\Delta T$ of VO$_2$ is 
greater than its critical temperature $T_c=340 K$ so that VO$_2$ is in its metallic phase 
while the temperature of the glass medium is $T_\rR=T_c-\Delta T<T_\rL$. 
The average temperature can be either positioned at $T_c$ (i.e. $\alpha=1$) or shifted to lower or higher values.  
(ii) In the reverse mode (R) $T_\rL = T_c-\Delta T$ so that VO$_2$ is in its crystalline phase and $T_\rR = T_c+\alpha\Delta T > T_\rL$. 
In its crystalline (monoclinic) phase, VO$_2$ behaves as an uniaxial medium.  Experimental datas show that~\cite{Baker} 
the optical axis of VO$_2$ films is orthogonal its surface [see Fig.~\ref{Fig:Perm} for a plot of the
permittivites parallel and perpendicular to the surface]. The net heat flux exchanged per unit surface between 
two isotropic media or between one uniaxial and one isotropic medium can be written in the general form~\cite{Polder1971,BiehsEtAl2010,JoulainPBA2010}
\begin{equation}
\begin{split}
  \Phi_{F/R} &= \int_0^\infty\!\frac{{\rm d}\omega}{2 \pi} \Delta\Theta(\omega)\!\! \sum_{j = \{\rm s,p\}}\int\! \frac{{\rm d}^2 \boldsymbol{\kappa}}{(2 \pi)^2} \, \mathcal{T}_{j,F/R}(\omega,\boldsymbol{\kappa}; d) \\
             &= \int_0^\infty\!{\rm d}\omega\, \Delta\Theta(\omega) \varphi_{F/R}(\omega,d) 
\end{split}
\label{Eq:SpectralPoynting}
\end{equation}
where $\Delta\Theta(\omega)=\Theta(\omega,T_L)-\Theta(\omega,T_R)$ is the difference of mean energies of 
Planck oscillators at frequency $\omega$ and at the temperatures of two intercating media. As 
for $\mathcal{T}_{j,F/R}(\omega,\boldsymbol\kappa)$ represents the energy transmission probability carried by 
the mode $(\omega,\boldsymbol{\kappa})$ ($\boldsymbol{\kappa}$ is the lateral wave vector) in one of two 
polarization states (s and p polarization). It is defined for the propagating modes with $\kappa < \omega/c$ by~\cite{BiehsEtAl2011}
\begin{equation}
   \mathcal{T}_{j,F/R}(\omega,\kappa; d) =
     \Tr[(\mathds{1}-\mathds{R}^{\dagger}_2\mathds{R}_2)\mathds{D}^{12}(\mathds{1}-\mathds{R}^{\dagger}_1\mathds{R}_1)\mathds{D}^{12\dagger}]
\label{Eq:TransmissionCoeff}
\end{equation}
where the reflection matrix of each interface is given by ($l = 1,2$) 
\begin{eqnarray}
\label{ReflectionMatrices}
\mathds{R}_l = \left[
\begin{array}{cc}
   r^{{\rm s, s}}_l (\omega, \kappa) &  r^{{\rm s, p}}_l (\omega, \kappa) \\
   r^{{\rm p, s}}_l (\omega, \kappa) &  r^{{\rm p, p}}_l (\omega, \kappa) 
\end{array} \right] ,
\end{eqnarray}
and the matrix $\mathds{D}^{12}$ is defined as
\begin{equation}
\mathds{D}^{12}= {(\mathds{1} - \mathds{R}_1\mathds{R}_2 e^{2 \ri k_{z0} d})}^{-1},
\end{equation}
with $k_{z0} = \sqrt{\omega^2/c^2 - \kappa^2}$. The matrix elements $r^{j,j'}_l$ 
of the reflection matrix are the reflection coefficients for the scattering of an incoming $j$-polarized plane
wave into an outgoing $j'$-polarized wave. For isotropic or uniaxial media with the optical axis orthogonal 
to the surface $r^{{\rm s, p}}_l=r^{{\rm p, s}}_l=0$. The remaining reflection coefficients are given by
\begin{equation}
  r^{{\rm s, s}}_l=\frac{k_{z0}-k_{l;s}}{k_{z0}+k_{l;s}},
  \label{Eq:r_ss}
\end{equation}
\begin{equation}
  r^{{\rm p, p}}_l=\frac{\epsilon_\parallel k_{z0}-k_{l;p}}{\epsilon_\parallel k_{z0}+k_{l;p}},
  \label{Eq:r_pp}
\end{equation}
where $k_{l;s,p}$ are solutions of the Fresnel equation
\begin{equation}
  \biggl(\epsilon_\parallel \frac{\omega^2}{c^2}-\kappa^2-k_{l;s}^2\biggl)\biggr(\epsilon_\parallel \epsilon_{\perp}\frac{\omega^2}{c^2}-\epsilon_\parallel \kappa^2-\epsilon_{\perp}k_{l;p}^2 \biggr)=0.
\label{Eq:Fresnel}
\end{equation}
Here $\epsilon_\parallel$ and $\epsilon_\perp$ are the permittivities parallel and perpendicular to the surface of
the uniaxial material VO$_2$. For amorphous glass which is isotropic we have 
$\epsilon_\parallel = \epsilon_\perp = \epsilon_{\text{SiO}_2}$ \cite{Palik}. Finally, we simplify the above expressions for the 
transmission coefficients by taking the limit $d \rightarrow \infty$ and obtain
\begin{equation}
   \mathcal{T}_{j,F/R}(\omega,\kappa) = \frac{(1 - |r^{{\rm j, j}}_1 (\omega, \kappa)|^2 )(1 - |r^{{\rm j, j}}_2 (\omega, \kappa)|^2)}{1 - |r^{{\rm j, j}}_1 (\omega, \kappa)|^2 |r^{{\rm j, j}}_2 (\omega, \kappa)|^2}.
\label{Eq:TransCoeff}
\end{equation}

\begin{figure}[Hhbt]
\includegraphics[scale=0.35]{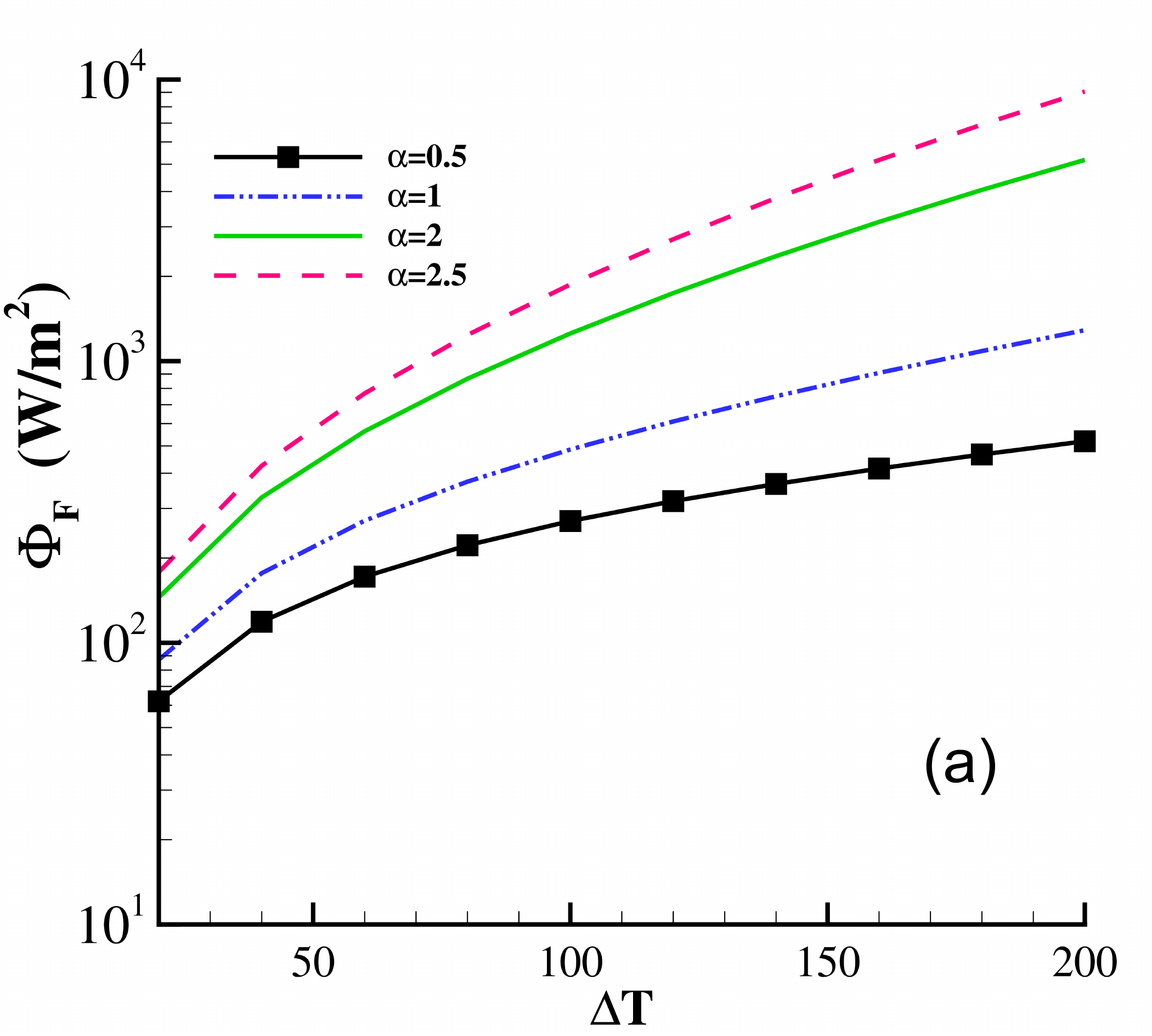}
\includegraphics[scale=0.35]{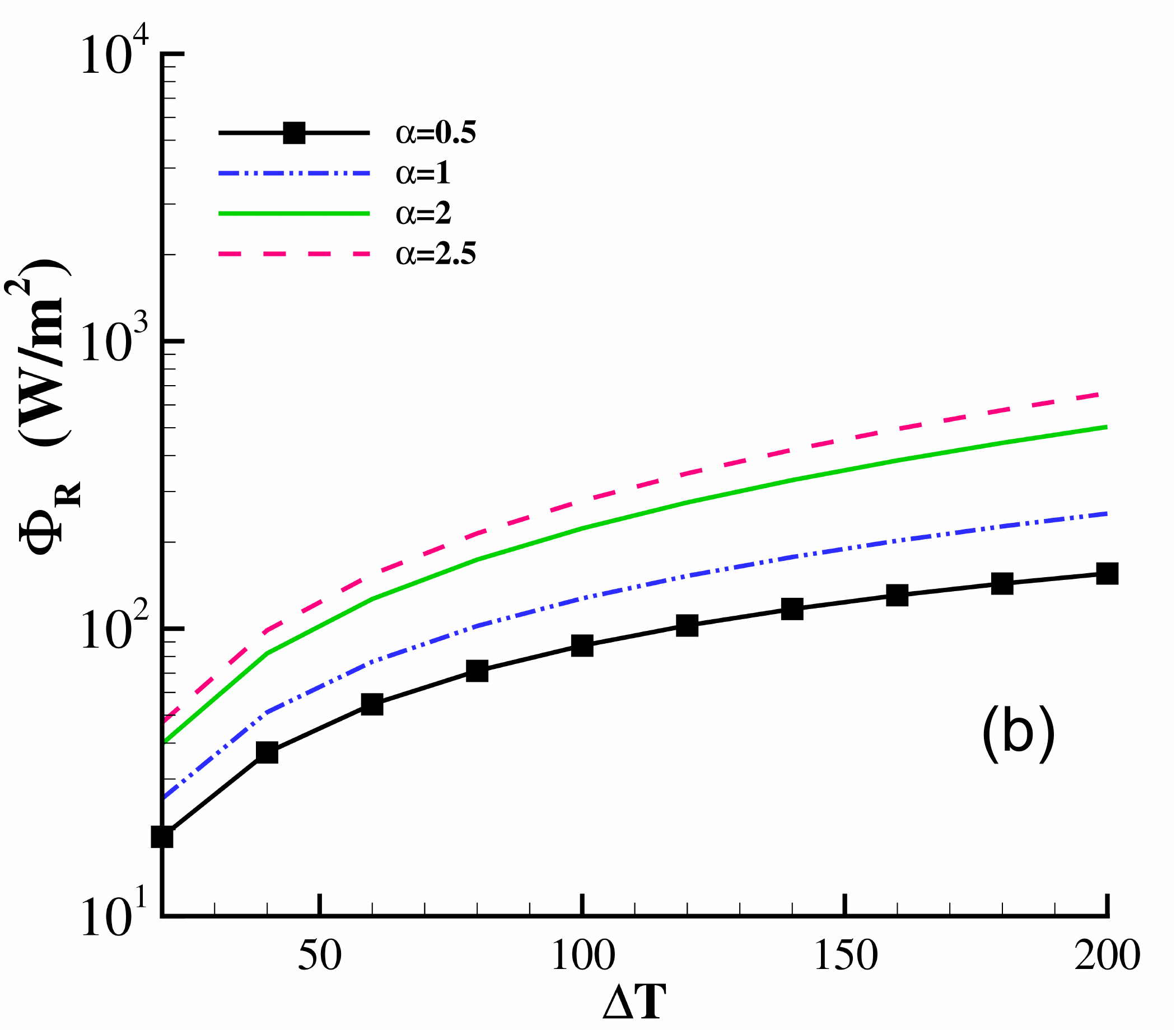}
\caption{Orientation dependence of heat flux in (a) the forward  and (b) in the reversed temperature scenario 
         between VO$_2$ and glass. The dielectric permittivities of glass and VO$_2$ are taken from Refs.~\cite{Palik} and~\cite{Baker} 
         while the critical transition temperature of VO$_2$ is $T_c=340\,{\rm K}$. \label{Fig:PhiFR}}
\end{figure}
\begin{figure}[Hhbt]
\includegraphics[scale=0.7]{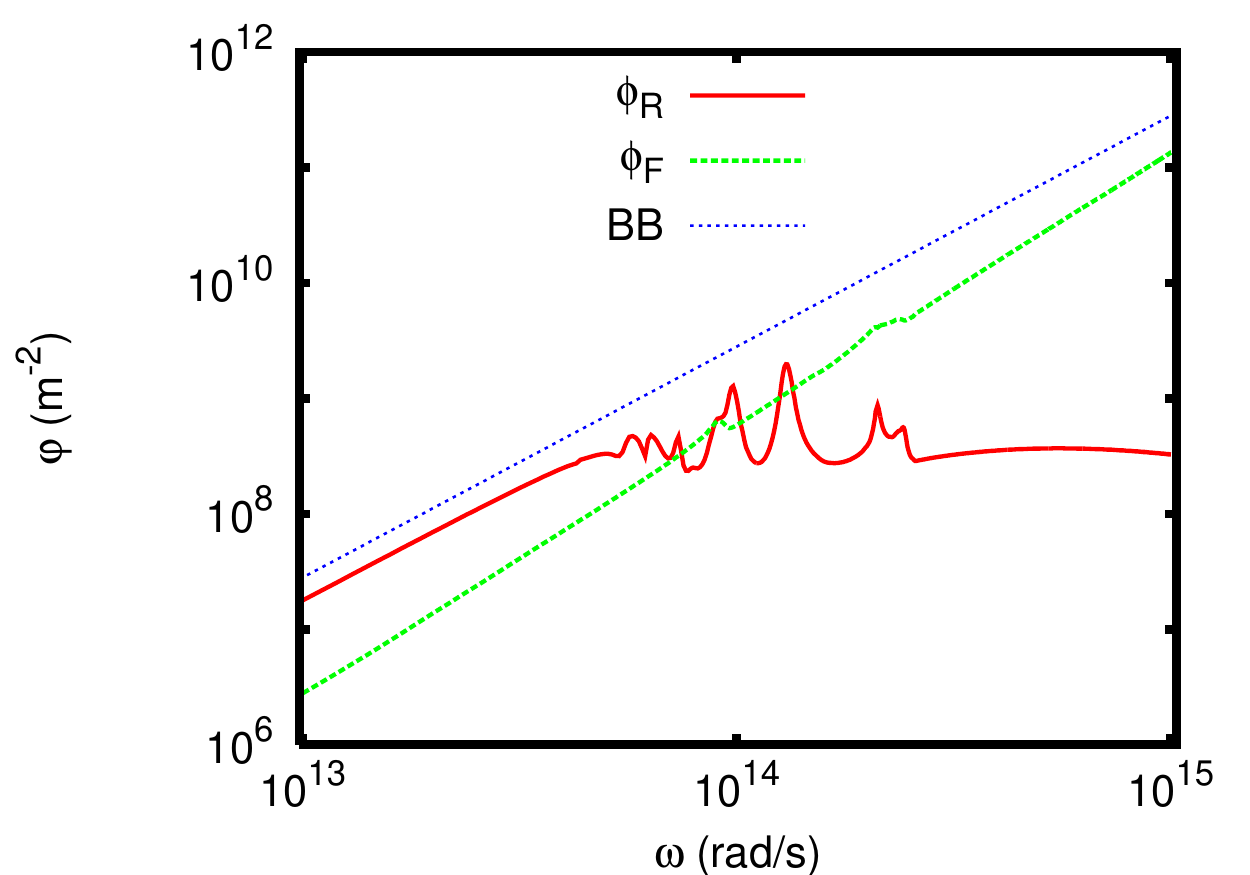}
\caption{Plot of the spectral heat flux $\varphi_{F/R}(\omega, d\rightarrow\infty)$ introduced in Eq.~(\ref{Eq:SpectralPoynting}) 
         as a function of frequency for forward direction where VO$_2$ is metallic
         and the reverse situation where VO$_2$ is crystalline. In addition, we have plotted 
         $\varphi = \frac{\omega^2}{\pi^2 c^3} \frac{c}{4}$ for the case that both materials are perfect black 
         bodies, i.e.\ $\mathcal{T}_{j,R/F} = 1$. \label{Eq:Spectrum}}
\end{figure}

Figures~\ref{Fig:PhiFR}(a) and \ref{Fig:PhiFR}(b) show the flux exchanged in the far field for the forward and the 
reversed scenario with respect to the temperture difference and the degree of asymetry of the temperature gradient. 
The evolution of fluxes $\Phi_F$ and $\Phi_R$ are very similar (they increase monoticaly with the temperature 
difference). In the forward scenario the magnitude of the flux is always more than 10 times greater than in 
the reverse scenario. In order to understand this difference we have plotted in Fig.~\ref{Eq:Spectrum} the
spectral heat flux $\varphi(\omega,d\rightarrow \infty)$ defined in Eq.~(\ref{Eq:SpectralPoynting}) 
using the transmission coefficient in Eq.~(\ref{Eq:TransCoeff}) for the forward and reversed scenario. 
With amorph VO$_2$, the spectral heat
flux is broadband and scales like $\propto \omega^2$ beyond $\omega=2.2\times10^{14}\,{\rm rad/s}$ that is in the 
frequency range where the diode operates when $T_\rL>T_c$. On the contrary, when VO$_2$ is crystalline, all 
propagating mode located at frequencies greater than $\omega=1.5\times10^{14}\,{\rm rad/s}$ give a very small and nearly 
constant contribution to the spectral heat flux and therefore do not transport much energy. This comes from 
the weak emissivity of VO$_2$ in this state. Hence, VO$_2$ changes for frequencies larger than approximately
$\omega=1.5\times10^{14}\,{\rm rad/s}$ from a metallic broadband emitter in the forward scenario to a very 
strong reflector (and hence very poor thermal emitter) in the reverse configuration. This asymetry is the key for  
obtaining a highly efficient phase-change thermal diode.

\begin{figure}[Hhbt]
\includegraphics[scale=0.35]{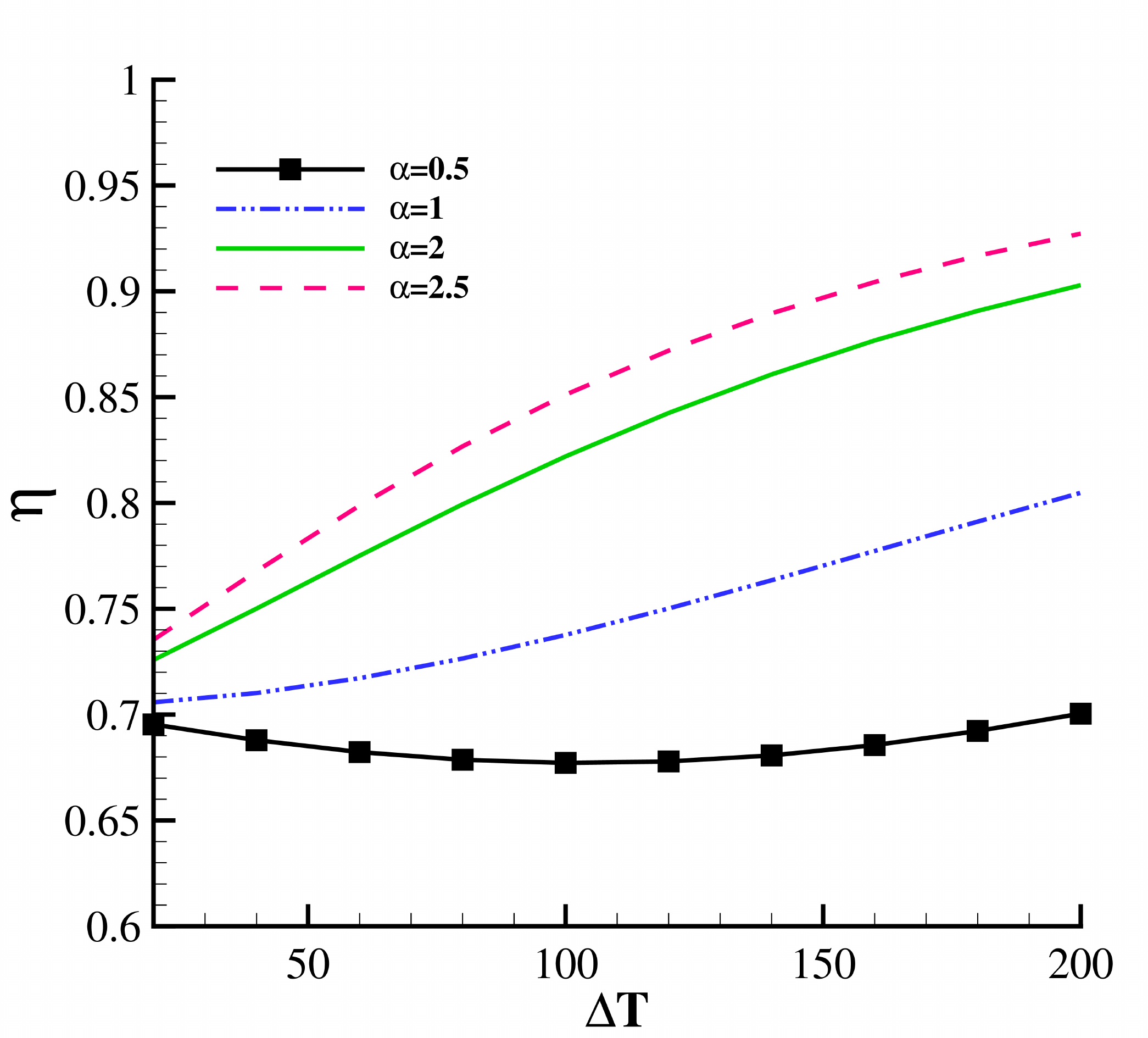}
     \caption{Rectification coefficient of a VO$_2$-glass system with respect to the temperature discrepancy 
              and the asymmetry of temperature gradient compared with the critical temperature $T_c$. \label{Fig:Eta}}
\end{figure}

The rectification coefficient $\eta$ of this rectifier is plotted in Fig.~\ref{Fig:Eta}. For very small temperature 
differences we find that $\eta\simeq70\%$ illustrating the high efficiency of the phase transition of VO$_2$. 
This coefficient grows to $\thicksim92\%$ for large temperature differences $\Delta T=200$ and for a large asymetry factor of $\alpha=2.5$.
On the contrary, when $\alpha$ is small the thermal rectification does not grow significantly with $\Delta T$. 
In such a configuration, the distribution function involves the modes of high frequencies in the forward scenario and the 
modes of low frequency in the reverse scenario. But, according to Fig. 3, the efficiency of heat transport by these modes 
is relatively high in both cases. As consequence, the flux exchanged in both scenarios are similar as we see in Fig. 2 

In conlcusion, we have introduced the concept of a phase-change radiative rectifier. Thanks to the bifurfaction in the 
optical behavior of MITs close to the critical temperature, large thermal rectification coefficients can been 
obtained. The here obtained rectification coefficients are between 70\% and 92\% which is rather high as compared
to previous radiative rectifier concepts which show rectification coefficients 
less than 44\%, 52\% or 70\%~\cite{OteyEtAl2010,IizukaFan2012,BasuFrancoeur2011,NefzaouiEtAl2013}. 
Hence, by means of using phase change materials, very efficient thermal diodes can be designed. 
Beyond their potential for thermal managment, these radiative thermal rectifiers suggest the possibility to 
develop out of contact thermal analogs of electronic devices such as radiative thermal transistors and radiative 
thermal memories for processing information by utilizing photons rather than electrons and thermal sources 
rather than electric currents.

%
%


\end{document}